\documentclass[preprint,showpacs,showkeys,preprintnumbers,amsmath,amssymb]{revtex4}

\usepackage{graphicx}
\usepackage{dcolumn}
\usepackage{bm}

\begin{document}

\title{Dual Bosonic Thermal Green Function and Fermion Correlators
of the Massive Thirring Model at a Finite Temperature}

\author{Leonardo Mondaini}
 \email{mondaini@cbpf.br}
\affiliation{Centro Brasileiro de Pesquisas F\'{i}sicas, Rua Dr.
Xavier Sigaud 150, Rio de Janeiro-RJ 22290-180, Brazil}

\author{E. C. Marino}
 \email{marino@if.ufrj.br}
\affiliation{Instituto de F\'{i}sica, Universidade Federal do Rio de
Janeiro, Rio de Janeiro-RJ 21941-972, Brazil}

\date{\today}

\begin{abstract}
The Euclidian thermal Green function of the two-dimensional (2D)
free massless scalar field in coordinate space is written as the
real part of a complex analytic function of a variable that
conformally maps the infinite strip $-\infty<x<\infty$
($0<\tau<\beta$) of the $z=x+i\tau$ ($\tau$: imaginary time) plane
into the upper-half-plane. Using this fact and the Cauchy-Riemann
conditions, we identify the dual thermal Green function as the
imaginary part of that function. Using both the thermal Green
function and its dual, we obtain an explicit series expression for
the fermionic correlation functions of the massive Thirring model
(MTM) at a finite temperature.
\end{abstract}



\maketitle


In recent publications, attention has been focused on the thermal
Green function of the free massless scalar field in 2D and its
relation with the bosonization of the Thirring model, either massive
or massless and using both imaginary-time \cite{delepine,nicola} and
real-time \cite{RLPG,belvedere} formalisms. In the former, the
thermal correlation functions of bilinears of the fermion field
(which are bosonic) of the MTM were derived within the bosonized
framework, whereas in the latter, the case of the massless Thirring
model is considered.

In the present work, using the imaginary-time formalism, we obtain
an explicit series expression for the fermionic thermal correlators
of the MTM, thereby filling a gap existing in the subject of
bosonization at a finite temperature. In order to do that, we
rewrite the thermal Green function in a way that allows us to easily
recognize it as the real part of an analytic function. This fact
leads us to determine the corresponding dual thermal Green function
as the imaginary part of that function, according to the
Cauchy-Riemann conditions. This dual thermal Green function turns
out to be a key ingredient for the obtainment of the
\emph{fermionic} correlators.

The MTM is described by the Lagrangian density
\begin{equation}
{\mathcal L}_{MTM} = i\bar\psi\not\!\partial\psi - M_0 \bar \psi
\psi - \frac{g}{2}(\bar \psi \gamma_\mu \psi)(\bar \psi \gamma^\mu
\psi) , \label{3.1}
\end{equation}
where $\psi$ is a two-component Dirac fermion field in
(1+1)-dimensions. It is well known that it can be mapped, both at
$T=0$ \cite{col} and $T\neq 0$ \cite{delepine,nicola}, into the
sine-Gordon (SG) theory of a scalar field, whose dynamics is
determined by
\begin{equation}
{\mathcal L}_{SG} = \frac{1}{2}\partial_\mu\phi\
\partial^\mu\phi +2\alpha_0 \cos\eta\phi ,
\label{3.2}
\end{equation}
where the couplings in the two models are related as
\begin{equation}
g = \pi \left (\frac{4\pi}{\eta^2} - 1 \right), \ \ \ \ \ \ \ \ \ \
\ \ M_0 \bar \psi \psi = - 2\alpha_0 \cos\eta\phi . \label{3.2a}
\end{equation}
Under this mapping, the two components of the fermion field may be
expressed in terms of the SG field as
\begin{equation}
\psi_{1}(\vec r) =  \sigma(\vec r)\mu(\vec r) , \ \ \ \ \ \ \ \ \ \
\ \ \psi_{2}(\vec r) = \sigma^\dagger(\vec r)\mu(\vec r),
\label{3.3}
\end{equation}
where $\sigma(\vec r)$ and $\mu(\vec r)$ are, respectively, order
and disorder fields, satisfying a dual algebra, which can be
introduced in the SG theory \cite{ms}. These are given by
\begin{equation}
\sigma(x,\tau) = \exp \left\{ i\  \frac{\eta}{2} \ \phi (x,\tau)
\right\}, \label{3.4}
\end{equation}
\begin{equation}
\mu(x,\tau) = \exp \left\{ i\ \frac{2\pi}{\eta} \int_{-\infty}^{x}
dz\, \dot{\phi} (z,\tau) \right\}. \label{3.5}
\end{equation}
Equation (\ref{3.3}) coincides with the bosonized expression for the
fermion field, first obtained in \cite{mand} for the $T=0$ case and
also shown to hold at finite temperature in \cite{RLPG,belvedere}.

The Euclidian vacuum functional of the SG theory, for an arbitrary
$T$ may be written as the grand-partition function of a classical 2D
gas of point charges $\pm \eta$, contained in an infinite strip of
width $\beta=1/k_BT$, interacting through the potential $G_T(\vec
r)$, namely \cite{nicola}
\begin{equation}
\begin{split}
\mathcal{Z}& =
\sum_{m=0}^{\infty}\frac{\alpha_0^m}{m!}\sum_{\{\lambda_i\}_m}\int_0^\beta\int_{-\infty}^{\infty}
\prod_{i=1}^{m} d\tau_i\, dz_i \\
& \quad
\times\exp\left\{-\frac{\eta^2}{2}\sum_{i=1}^{m}\lambda_i\sum_{j=1}^{m}\lambda_j
G_T(\vec z_i-\vec z_j)\right\},
\end{split}
\label{3.6.1}
\end{equation}
where $\lambda_i=\pm 1$, $\sum_{\{\lambda_i\}_m}$ runs over all
possibilities in the set $\{\lambda_1,\ldots,\lambda_m\}$, and
$G_T(\vec r)$ is the thermal Euclidian Green function of the 2D free
massless scalar theory in coordinate space ($\vec r\equiv
(x,\tau)$), which is given by \cite{Das}
\begin{equation}
G_T(\vec
r)=\frac{1}{\beta}\sum_{n=-\infty}^{\infty}\int_{-\infty}^{\infty}
\frac{dk}{2\pi}\frac{e^{-i(kx+\omega_n\tau)}}{k^2+\omega_n^2},
\label{gft1}
\end{equation}
with $\omega_n=2\pi n/\beta$. This has been evaluated in
\cite{delepine} and is given by
\begin{equation}
G_T(\vec
r)=-\frac{1}{4\pi}\ln\left\{\frac{\mu_0^2\,\beta^2}{\pi^2}\left[\cosh\left(\frac{2\pi}
{\beta}x\right)-\cos\left(\frac{2\pi}{\beta}\tau\right)\right]\right\}.
\label{greenfunction}
\end{equation}
This has also been obtained by using methods of integration on the
complex plane \cite{LM}. At $T=0$, $G_T(\vec r)$ reduces to the 2D
Coulomb potential and we retrieve the usual mapping onto the Coulomb
gas \cite{sgcg}.

We may rewrite the above thermal Green function in terms of the new
complex variable
\begin{equation}
\zeta(\vec r)\equiv
\zeta(z)=\frac{\beta}{\pi}\sinh\left(\frac{\pi}{\beta}\,z\right),
\label{zeta}
\end{equation}
where  $z=x+i\tau$, as
\begin{equation}
G_T(\vec r)=\lim_{\mu_0\rightarrow
0}-\frac{1}{4\pi}\ln\left[\mu_0^2\zeta(\vec r)\zeta^*(\vec
r)\right]. \label{gft20}
\end{equation}

 At this point let us make a few comments about
(\ref{gft20}). Firstly, we note that in the zero temperature limit
($T\rightarrow 0$, $\beta\rightarrow \infty$), we have $\zeta
(z)\rightarrow z $ and $\zeta^*(z)\rightarrow z^* $ and, therefore,
we recover the well-known Green function at zero temperature, namely
\begin{equation}
\lim_{\beta\rightarrow\infty}G_T(\vec r;\mu_0)=-\frac{1}{4\pi} \ln
\left[\mu_0^2\, zz^*\right]=-\frac{1}{4\pi} \ln \left[\mu_0^2 |\vec
r|^2 \right]. \label{gft22}
\end{equation}

Comparing (\ref{gft20}) with (\ref{gft22}) we can see that the only
effect of a finite temperature is to exchange the complex variable
$z$ for $\zeta(z)$. Since $\zeta(z)$ is analytic, we conclude that
the thermal Green function is obtained from the one at zero
temperature by the following conformal mapping \cite{belavin}: the
infinite strip $0<\tau<\beta$ and $-\infty<x<\infty$ is mapped into
the region within the upper-half-$\zeta$-plane.

From Eq. (\ref{gft20}) we can also see that the thermal Green
function may be written as the real part of an analytic function of
the complex variable $\zeta$ namely
\begin{equation}
G_T(\vec
r;\mu_0)=\textrm{Re}\left[\mathcal{F}(\zeta)\right]=\frac{1}{2}\left[\mathcal{F}(\zeta)+\mathcal{F}^*(\zeta)\right],
\label{gft20b}
\end{equation}
where $\mathcal{F}(\zeta)\equiv -(1/2\pi)\ln\left[\mu_0\zeta(\vec
r)\right]$.

The imaginary part of $\mathcal{F}(\zeta)$ may be written as
\begin{equation}
\begin{split}
\tilde{G}_T(\vec
r)\equiv\textrm{Im}\left[\mathcal{F}(\zeta)\right]&=\frac{1}{2i}
\left[\mathcal{F}(\zeta)-\mathcal{F}^*(\zeta)\right]\\&=-\frac{1}{4\pi
i}\ln\left[\frac{\zeta(\vec r)}{\zeta^*(\vec r)}\right].
\label{gft20c}
\end{split}
\end{equation}

Now, from the analyticity of $\mathcal{F}(\zeta)$, then, it follows
that its imaginary and real parts must satisfy the Cauchy-Riemann
conditions, which are given by
\begin{equation}
\epsilon^{\mu\nu}\partial_\nu G_T =-\partial_\mu \tilde{G}_T, \ \ \
\ \ \ \ \ \ \ \ \ \epsilon^{\mu\nu}\partial_\nu
\tilde{G}_T=\partial_\mu G_T. \label{cauchyriemann}
\end{equation}
This property characterizes $\tilde{G}_T$ as the dual thermal Green
function.

Returning to the Euclidian vacuum functional of the SG theory, we
see that this may be now written as
\begin{equation}
\begin{split}
\mathcal{Z}&=\lim_{\varepsilon\rightarrow 0}\lim_{\mu_0\rightarrow
0}
\sum_{n=0}^{\infty}\frac{\alpha^{2n}}{(n!)^2}\int_0^\beta\int_{-\infty}^{\infty}
\prod_{i=1}^{2n} d\tau_i\, dz_i
\\ & \quad \times\exp\left\{\frac{\eta^2}{8\pi}\sum_{i\neq j=1}^{2n}
\lambda_i\lambda_j\ln \left\{\mu^2_0 \left[\zeta(\vec z_i-\vec
z_j)\zeta^*(\vec z_i-\vec z_j)
+|\varepsilon|^2\right]\right\}\right\}, \label{3.6b}
\end{split}
\end{equation}
where, in order to obtain (\ref{3.6b}), we have used the
UV-regulated version of $G_T(\vec r;\mu_0)$, namely
\begin{equation}
G_T(\vec r;\mu_0,\varepsilon) = -\frac{1}{4\pi} \ln \left\{\mu^2_0
\left[\zeta(\vec r)\zeta^*(\vec r)
+|\varepsilon|^2\right]\right\}.\label{3.7a}
\end{equation}
The renormalized coupling $\alpha$ is related to the one in
(\ref{3.2}) by $\alpha = \alpha_0
\left(\mu_0^2|\varepsilon|^2\right)^{\eta^2/(8\pi)}$ \cite{col}.

As in the $T=0$ case, existence of the $\mu_0\rightarrow 0$ limit
imposes the neutrality of the new gas, namely $\sum_{i=1}^m
\lambda_i=0$, because in this case the $\mu_0$-factors are
completely canceled. This implies that the index $m$ appearing in
(\ref{3.6.1}) must be even ($m=2n$) and, therefore,
$\sum_{\{\lambda_i\}_m}=(2n)!/(n!)^2$.

We can now determine the fermionic correlators at $T\neq 0$, by
using (\ref{3.3} -- \ref{3.5}) along with the gas representation of
the vacuum functional, just derived. The bosonization formulae
(\ref{3.3}) were shown to hold for $T\neq 0$ in the massless theory.
Notice however that, since we are making a mass expansion
($\alpha$-expansion), we are actually considering a sum of fermion
correlators in the massless theory. The insertion of $\sigma$ and
$\mu$ operators corresponds, respectively, to the introduction of
external charges (of magnitude $\eta/2$) and ``magnetic" fluxes on
the gas \cite{ms}. The fermion correlators, then are nothing but the
exponential of the interaction energy of the associated classical
system. Charges and ``magnetic" fluxes interact with their similar,
through the thermal Green function $G_T(\vec r)$, whereas the
charge-flux interaction occurs via the dual thermal Green function
$\tilde{G}_T(\vec r)$ \cite{ms}. This is the reason why it is
crucial to know this function in order to obtain the fermion
correlators. In the case of (charge conserving) fermion bilinear
correlators, only external charges are inserted into the gas and,
therefore, only the thermal Green function $G_T(\vec r)$ is
required.

Following the above considerations and the same procedure employed
at $T=0$ \cite{MM2}, we can write  the four
 components of the two-point fermion correlation function of the
  MTM at finite temperature as
\begin{equation}
\begin{split}
\langle&\psi_{1(2)}(\vec x)\psi_{1(2)}^\dagger(\vec y)\rangle
\\ &=
\mathcal{Z}^{-1}\sum_{m=0}^{\infty}\frac{\alpha_0^m}{m!}\sum_{\{\lambda_i\}_m}\int_0^\beta\int_{-\infty}^{\infty}
\prod_{i=1}^{m} d\tau_i\, dz_i\,\exp \left\{\frac{1}{2}\int d^2z\,
d^2z' \right . \\ & \quad \left . \times
\left[\left(i\eta\sum_{i=1}^m \lambda_i\,\delta^2(\vec z-\vec
{z_i})+\frac{2\pi}{\eta}\left[\int_{\vec x}^{\vec y}d\eta_\mu\,
\epsilon^{\alpha\mu}\delta^2(\vec z - \vec
\eta)\right]\partial_\alpha^{(z)}\right .\right .\right .\\ & \quad
\left .\left .\left . +(-) i\,\frac{\eta}{2}\left[\delta^2(\vec
z-\vec x)-\delta^2(\vec z-\vec
y)\right]\right)\left(i\eta\sum_{j=1}^m \lambda_j\,\delta^2(\vec
{z'}-\vec {z_j})\right .\right .\right .\\ & \quad  \left .\left
.\left . +\frac{2\pi}{\eta}\left[\int_{\vec x}^{\vec y}d\xi_\nu\,
\epsilon^{\beta\nu}\delta^2(\vec {z'} - \vec
\xi)\right]\partial_\beta^{(z')}+(-)
i\,\frac{\eta}{2}\left[\delta^2(\vec {z'}-\vec x)-\delta^2(\vec
{z'}-\vec y)\right]\right)\right .\right .\\&\quad \left .\left
.\times \,G_T(\vec z-\vec {z'})\right]\right\}
\end{split}
\label{e3}
\end{equation}
and
\begin{equation}
\begin{split}
\langle&\psi_{1(2)}(\vec x)\psi_{2(1)}^\dagger(\vec y)\rangle
\\ &=
\mathcal{Z}^{-1}\sum_{m=0}^{\infty}\frac{\alpha_0^m}{m!}\sum_{\{\lambda_i\}_m}\int_0^\beta\int_{-\infty}^{\infty}
\prod_{i=1}^{m} d\tau_i\, dz_i\,\exp \left\{\frac{1}{2}\int d^2z\,
d^2z' \right .\\ & \quad \left .\times \left[\left(i\eta\sum_{i=1}^m
\lambda_i\,\delta^2(\vec z-\vec
{z_i})+\frac{2\pi}{\eta}\left[\int_{\vec x}^{\vec y}d\eta_\mu\,
\epsilon^{\alpha\mu}\delta^2(\vec z - \vec
\eta)\right]\partial_\alpha^{(z)}\right .\right .\right .\\ & \quad
\left .\left .\left . +(-) i\,\frac{\eta}{2}\left[\delta^2(\vec
z-\vec x)+\delta^2(\vec z-\vec
y)\right]\right)\left(i\eta\sum_{j=1}^m
\lambda_j\,\delta^2(\vec {z'}-\vec {z_j})\right .\right .\right .\\
& \quad  \left .\left .\left . +\frac{2\pi}{\eta}\left[\int_{\vec
x}^{\vec y}d\xi_\nu\, \epsilon^{\beta\nu}\delta^2(\vec {z'} - \vec
\xi)\right]\partial_\beta^{(z')}+(-)
i\,\frac{\eta}{2}\left[\delta^2(\vec {z'}-\vec x)+\delta^2(\vec
{z'}-\vec y)\right]\right)\right .\right .\\&\quad \left .\left
.\times \,G_T(\vec z-\vec {z'})\right]\right\}.
\end{split}
\label{e3}
\end{equation}
After some algebra, we get
\begin{equation}
\begin{split}
\langle&\psi_{1(2)}(\vec x)\psi_{1(2)}^\dagger(\vec y)\rangle
\\ &=
\mathcal{Z}^{-1}\sum_{m=0}^{\infty}\frac{\alpha_0^m}{m!}\sum_{\{\lambda_i\}_m}\int_0^\beta\int_{-\infty}^{\infty}
\prod_{i=1}^{m} d\tau_i\, dz_i \,\exp
\left\{\frac{\eta^2}{8\pi}\sum_{i\neq j=1}^{m}\lambda_i \lambda_j
G_T(\vec z_i-\vec z_j)\right .
\\ & \quad \left .+(-)\frac{\eta^2}{8\pi}\sum_{i=1}^{m}\lambda_i\, \left[G_T(\vec
z_i-\vec x)-G_T(\vec z_i-\vec y)\right]+\frac{\eta^2}{16\pi}
\left[G_T(\vec 0)-G_T(\vec x-\vec y)\right]\right .
\\ & \quad \left .-\frac{i}{2}\sum_{i=1}^{m}\lambda_i\int_{\vec x}^{\vec y}d\xi_\nu
\epsilon^{\beta\nu}\partial_\beta^{(\xi)}G_T(\vec z_i-\vec
\xi)-\frac{\pi}{2\eta^2}\int_{\vec x}^{\vec y}d\eta_\mu\int_{\vec
x}^{\vec y}d\xi_\nu\,
\epsilon^{\alpha\mu}\partial_\alpha^{(\eta)}\epsilon^{\beta\nu}
\partial_\beta^{(\xi)}\right . \\ & \quad \left . \times \,G_T(\vec \eta-\vec \xi)-(+)\,\frac{i}{4}\int_{\vec x}^{\vec y}d\xi_\nu
\epsilon^{\beta\nu}\partial_\beta^{(\xi)}\left[G_T(\vec x-\vec
\xi)-G_T(\vec y-\vec \xi)\right]\right\}
\end{split}
\label{e5}
\end{equation}
and
\begin{equation}
\begin{split}
\langle&\psi_{1(2)}(\vec x)\psi_{2(1)}^\dagger(\vec y)\rangle
\\ &=
\mathcal{Z}^{-1}\sum_{m=0}^{\infty}\frac{\alpha_0^m}{m!}\sum_{\{\lambda_i\}_m}\int_0^\beta\int_{-\infty}^{\infty}
\prod_{i=1}^{m} d\tau_i\, dz_i \,\exp
\left\{\frac{\eta^2}{8\pi}\sum_{i\neq j=1}^{m}\lambda_i \lambda_j
G_T(\vec z_i-\vec z_j)\right .
\\ & \quad \left .+(-)\frac{\eta^2}{8\pi}\sum_{i=1}^{m}\lambda_i\, \left[G_T(\vec
z_i-\vec x)+G_T(\vec z_i-\vec y)\right]+\frac{\eta^2}{16\pi}
\left[G_T(\vec 0)+G_T(\vec x-\vec y)\right]\right .
\\ & \quad \left .-\frac{i}{2}\sum_{i=1}^{m}\lambda_i\int_{\vec x}^{\vec y}d\xi_\nu
\epsilon^{\beta\nu}\partial_\beta^{(\xi)}G_T(\vec z_i-\vec
\xi)-\frac{\pi}{2\eta^2}\int_{\vec x}^{\vec y}d\eta_\mu\int_{\vec
x}^{\vec y}d\xi_\nu\,
\epsilon^{\alpha\mu}\partial_\alpha^{(\eta)}\epsilon^{\beta\nu}
\partial_\beta^{(\xi)}\right . \\ & \quad \left . \times \,G_T(\vec \eta-\vec \xi)-(+)\,\frac{i}{4}\int_{\vec x}^{\vec y}d\xi_\nu
\epsilon^{\beta\nu}\partial_\beta^{(\xi)}\left[G_T(\vec x-\vec
\xi)+G_T(\vec y-\vec \xi)\right]\right\}.
\end{split}
\label{e5}
\end{equation}
Finally, using (\ref{3.7a}), (\ref{gft20c}), and
(\ref{cauchyriemann}), we obtain
\begin{equation}
\begin{split}
\langle&\psi_{1(2)}(\vec x)\psi_{1(2)}^\dagger(\vec y)\rangle \\
&=\lim_{\varepsilon\rightarrow 0}\lim_{\mu_0\rightarrow 0}
{\mathcal{Z}}^{-1} \left[\frac{\zeta(\vec x-\vec y)\zeta(\vec y-\vec
x)}{\zeta^*(\vec x-\vec y)\zeta^*(\vec y-\vec
x)}\right]^{+(-)\frac{1}{4}}\left[ \frac{|\varepsilon|^2}{\zeta(\vec
x-\vec y)\zeta^*(\vec x-\vec
y)}\right]^{(\frac{\pi}{\eta^2}+\frac{\eta^2}{16\pi})}
\\ & \quad \times \sum_{n=0}^{\infty}\frac{\alpha^{2n}}{(n!)^2}\int_0^\beta\int_{-\infty}^{\infty}
\prod_{i=1}^{2n} d\tau_i\, dz_i \\ & \quad \times \exp \left
\{\frac{\eta^2}{8\pi}\sum_{i\neq j=1}^{2n}\lambda_i\lambda_j\ln
\left\{\mu^2_0 \left[\zeta(\vec z_i-\vec z_j)\zeta^*(\vec z_i-\vec
z_j) +|\varepsilon|^2\right]\right\}\right . \\ & \quad \left .+(-)
\frac{\eta^2}{8\pi}\sum_{i=1}^{2n}\lambda_i\ln
\frac{\left[\zeta(\vec z_i-\vec x)\zeta^*(\vec z_i-\vec x) +
|\varepsilon|^2\right]}{\left[\zeta(\vec z_i-\vec y)\zeta^*(\vec
z_i-\vec y) + |\varepsilon|^2\right]}+\frac{1}{2}\sum_{i=1}^{2n}
\lambda_i\ln \frac{\zeta(\vec z_i-\vec y)\zeta^*(\vec z_i-\vec
x)}{\zeta^*(\vec z_i-\vec y)\zeta(\vec z_i-\vec x)} \right \}
\end{split}
\label{3.8}
\end{equation}
and
\begin{equation}
\begin{split}
\langle&\psi_{1(2)}(\vec x)\psi_{2(1)}^\dagger(\vec y)\rangle \\
&=\lim_{\varepsilon\rightarrow 0}\lim_{\mu_0\rightarrow 0}
{\mathcal{Z}}^{-1}\left[\frac{\zeta(\vec x-\vec y)\zeta^*(\vec
y-\vec x)}{\zeta^*(\vec x-\vec y)\zeta(\vec y-\vec
x)}\right]^{+(-)\frac{1}{4}}\left[\mu_0^2|\varepsilon|^2\right]^{(\frac{\pi}{\eta^2}
+\frac{\eta^2}{16\pi})}
 \\ & \quad
\times \left\{\mu_0^2\left[\zeta(\vec x-\vec y)\zeta^*(\vec x-\vec
y)\right]\right\}^{-(\frac{\pi}{\eta^2}-\frac{\eta^2}{16\pi})}
\sum_{n=0}^{\infty}\frac{\alpha^{(2n+1)}}{n!(n+1)!}\int_0^\beta\int_{-\infty}^{\infty}
\prod_{i=1}^{2n+1} d\tau_i\, dz_i \\ & \quad \times \exp \left
\{\frac{\eta^2}{8\pi}\sum_{i\neq j=1}^{2n+1}\lambda_i\lambda_j\ln
\left\{\mu^2_0 \left[\zeta(\vec z_i-\vec z_j)\zeta^*(\vec z_i-\vec
z_j)
+|\varepsilon|^2\right]\right\}+(-)\frac{\eta^2}{8\pi}\sum_{i=1}^{2n+1}\lambda_i
\right . \\ & \quad \left .\times\ln
\left\{\left\{\mu_0^2\left[\zeta(\vec z_i-\vec x)\zeta^*(\vec
z_i-\vec x) + |\varepsilon|^2\right]\right\}
\left\{\mu_0^2\left[\zeta(\vec z_i-\vec y)\zeta^*(\vec z_i-\vec y) +
|\varepsilon|^2\right]\right\}\right\} \right .\\ & \quad \left
.+\frac{1}{2}\sum_{i=1}^{2n+1} \lambda_i\ln \frac{\zeta(\vec
z_i-\vec y)\zeta^*(\vec z_i-\vec x)}{\zeta^*(\vec z_i-\vec
y)\zeta(\vec z_i-\vec x)} \right \} . \label{3.9}
\end{split}
\end{equation}
In (\ref{3.8}), we still have neutrality of the gas with $n$
positive and $n$ negative $\lambda_i$'s. In (\ref{3.9}), on the
other hand, we have $n$ positive and $n+1$ negative $\lambda_i$'s
for $\langle\psi_{1}\psi_{2}^\dagger\rangle$. Conversely, for
$\langle\psi_{2}\psi_{1}^\dagger\rangle$, we have $n$ negative and
$n+1$ positive $\lambda_i$'s. Notice, however, that overall
neutrality is still preserved in all functions if we consider the
external charges.

Let us finally remark that in the $T\rightarrow 0$ limit the above
correlators reduce to the corresponding functions of the zero
temperature theory \cite{MM2}. Also, observe that in the limit
$\alpha \rightarrow 0$ we recover the thermal fermion correlators of
the massless Thirring model. These are given by the pre-factor
multiplying the sum in (\ref{3.8}). In the case of the chirality
violating functions (\ref{3.9}), the massless limit yields zero as
it should.

\section*{Acknowledgments}

This work has been supported in part by CNPq and FAPERJ. LM was
supported by CNPq and ECM was partially supported by CNPq.


\begin{thebibliography}{20}
%
%
\bibitem{delepine} D. Del\'epine, R. Gonz\'alez Felipe, and J. Weyers, {\it Phys. Lett. B} \textbf{419},
296 (1998).
%
%
\bibitem{nicola} A. G\'omez Nicola and D. A. Steer, {\it Nucl. Phys.
B} \textbf{549}, 409 (1999).
%
%
\bibitem{RLPG} R. L. P. G. Amaral, L. V. Belvedere, and K. D. Rothe, {\it Ann. Phys. (N.Y.)} \textbf{320}, 399
(2005).
%
\bibitem{belvedere} L. V. Belvedere, {\it Phys. Rev. D} \textbf{74}, 107701 (2006).
%
%
\bibitem{col} S. Coleman, {\it Phys. Rev. D} \textbf{11}, 2088 (1975).
%
%
\bibitem{ms} E. C. Marino and J. A. Swieca, {\it Nucl. Phys. B} \textbf{170[FS1]}, 175
(1980).
%
%
\bibitem{mand} S. Mandelstam, {\it Phys. Rev. D} \textbf{11}, 3026 (1975).
%
%
\bibitem{Das} A. Das, {\it Finite Temperature Field Theory} (World
Scientific, 1997).
%
%
\bibitem{sgcg} S. Samuel, {\it Phys. Rev. D} \textbf{18}, 1916 (1978).
%
%
\bibitem{LM} L. Mondaini, Ph.D. thesis, IF-UFRJ,
2006.
%
%
\bibitem{belavin} A. A. Belavin, A. M. Polyakov, and A. B. Zamolodchikov, {\it Nucl. Phys. B} \textbf{241},
333 (1984).
%
%
\bibitem{MM2} L. Mondaini and E. C. Marino, {\it J. Phys. A: Math. Gen.} \textbf{39}, 967 (2006).
%
%
\end{thebibliography}
\end{document}